**Degenerate monolayer Ising superconductors via chiral-achiral molecule intercalation**

*Daniel Margineda\*, Covadonga Álvarez-García, Daniel Tezze, Sanaz Gerivani, Mohammad Furqan, Iván Rivilla, Fèlix Casanova, Raul Arenal, Emilio Artacho, Luis E. Hueso, and Marco Gobbi\**

D. Margineda, C. Álvarez-García, D. Tezze, S. Gerivani, F. Casanova, E. Artacho, L. E. Hueso
CIC nanoGUNE BRTA, Donostia-San Sebastian, Spain
E-mail: d.margineda@nanogune.eu

C. Álvarez-García, D. Tezze
Departamento de Polímeros y Materiales Avanzados: Física, Química y Tecnología, Facultad de Químicas (EHU), Apartado 1072, 20080, San Sebastián, Spain

S. Gerivani
Universidad del País Vasco (EHU), 48080, Bilbao, Spain

M. Furqan, R. Arenal
Instituto de Nanociencia y Materiales de Aragón (INMA) CSIC-Universidad de Zaragoza, 50009, Zaragoza, Spain

M. Furqan, R. Arenal
Laboratorio de Microscopías Avanzadas (LMA), Universidad de Zaragoza, Zaragoza, Spain

R. Arenal
ARAID Foundation, 50018 Zaragoza, Spain

I. Rivilla, E. Artacho
Donostia International Physics Center (DIPC), San Sebastian, 20018, Spain

I. Rivilla, F. Casanova, E. Artacho, L. E. Hueso, M. Gobbi
Ikerbasque, Basque Foundation for Science, 48009 Bilbao, Spain

E. Artacho

Theory of Condensed Matter, Cavendish Laboratory, University of Cambridge, J J Thomson Avenue, Cambridge CB3 0HE, United Kingdom

M. Gobbi

Centro de Física de Materiales – Materials Physics Center (CFM-MPC) CSIC-EHU, San Sebastian, 20018, Spain

E-mail: marco.gobbi@ehu.eus

**Abstract.** Engineering unconventional superconductors is a central challenge in condensed matter physics. Molecule-intercalated $TaS_2$ superlattices have recently been reported to host such states, yet their origin remains debated – underscoring the urgent need for controlled, device-integrated studies. Here, we report that nanometer-thick $TaS_2$ and $NbSe_2$ intercalated with chiral and achiral organic cations instead exhibit robust monolayer-like Ising superconductivity, with no evidence of unconventional pairing. Using high-quality superlattices integrated into devices, we disentangle the roles of interlayer coupling and charge transfer in shaping their superconducting behavior. In $TaS_2$, intercalation induces interlayer decoupling regardless of molecular size or symmetry, yielding monolayer-like Ising superconductivity. $NbSe_2$ instead retains quasi-three-dimensional transport, with a gradual Ising enhancement and near-monolayer behavior only at the largest interlayer spacing. Transport remains reciprocal across all superlattices, consistent with preserved inversion symmetry and incompatible with parity-breaking superconductivity and noncentrosymmetric monolayers. We attribute the behavior to electronically detached monolayers with opposite spin-split bands, coupled through thermal and tunneling processes, which overall preserve inversion symmetry. These findings establish molecular intercalation compounds as a robust, device-ready, platform for engineering advanced superconducting superlattices.

## 1. Introduction

Transition-metal dichalcogenides (TMDs) offer a fertile ground to explore unconventional superconductivity arising from the interplay of electronic correlations, spin-orbit coupling (SOC), and multiband Fermi surfaces[1–4]. In their 2H polyforms, metallic $NbSe_2$ and $TaS_2$ TMDs host conventional Bardeen-Cooper-Schrieffer (BCS) superconductivity [5, 6] coexisting with charge-density wave (CDW). These competing orders are strongly influenced by dimensionality [7, 8] and carrier density [9, 10]. In the two-dimensional (2D) limit, weakened interlayer coupling enables Ising SOC, which locks spins out of plane and drives in-plane critical fields far beyond the Pauli limit [11–13]. However, atomically thin TMDs have revealed unconventional states, including finite angular momentum states [14–16] and nodal pairing symmetries [17-19], challenging the long-standing assumption of nodeless, phonon-mediated BCS superconductivity. Yet, since TMDs are highly sensitive to extrinsic disorder [20] and unconventional states are easily suppressed by impurities [18, 21, 22], accessing their intrinsic superconducting properties requires high-quality, device-integrated monolayers. This motivates designing different strategies to decouple layers and stabilize unconventional behavior in more controllable systems [23-25].

Molecular intercalation offers an alternative route to engineer monolayer-like states by expanding the van der Waals gap and weakening interlayer coupling [26-29]. In $NbSe_2$, intercalation induces Ising enhancement [30], but to a lesser extent than in monolayers. Moreover, the transition temperature - drastically suppressed in $NbSe_2$ monolayers - remains almost unaffected upon intercalation, raising the question of whether true monolayer properties are realized.

In contrast, recent studies report unconventional superconductivity in $TaS_2$ superlattices [31-33], including through chiral-molecule intercalation [31,34]. Whether these states originate from molecular chirality [35], or from interlayer decoupling that stabilizes monolayer-like unconventional superconductivity remains unresolved.

Here, we demonstrate monolayer-like Ising superconductivity in nanometer-thick $TaS_2$ and $NbSe_2$ crystals intercalated with organic ions, including a chiral proline derivative, enabled by an in-situ galvanic intercalation method that is spontaneous and device-compatible. The evolution from bulk to Ising superconductivity is strikingly different in the two families. In $TaS_2$, intercalation induces interlayer decoupling for any molecular size, yielding clear hallmarks of monolayer-like transport. In $NbSe_2$, by contrast, only large intercalants and charge transfer drive the crossover from quasi-3D to 2D behavior, reflecting the stronger transverse dispersion across the Se orbitals. Importantly, dissipationless transport remains reciprocal, consistent with inversion-symmetric superlattices, incompatible with parity-breaking states observed in noncentrosymmetric 2H-TaS$_2$ monolayers. This suggests that molecular chirality alone is not sufficient to drive unconventional superconductivity. Instead, our results point to few-layer transport with spin-layer locking in electronically decoupled monolayers. In this picture, opposite spin polarization in adjacent layers – accessed thermally or through Josephson coupling preserves – global inversion symmetry.

## 2. Results

### 2.1. Spontaneous galvanic intercalation of thin flakes.

Thin flakes of TaS$_2$ and NbSe$_2$, with thickness ranging from 15 to 30 nm are in-situ intercalated with organic cations using a galvanic technique illustrated in **Figure 1**a,b (more details in methods and Ref. [36]). In this process, a few-nm-thick flake previously transferred onto prepatterned electrodes is placed in contact with a low-reduction potential metal (M$^0$) and then immersed in a solution of the target guest species. The galvanic intercalation is enabled if the reduction potential $E_{red}$ of the M$^0$ lies below the intercalation potential $E_{int}$ [36] as represented in Figure 1c. We found homogeneous intercalation of TaS$_2$ crystals with an indium anode, whereas NbSe$_2$ requires a metal with a larger $E_{red}$ offered by magnesium.

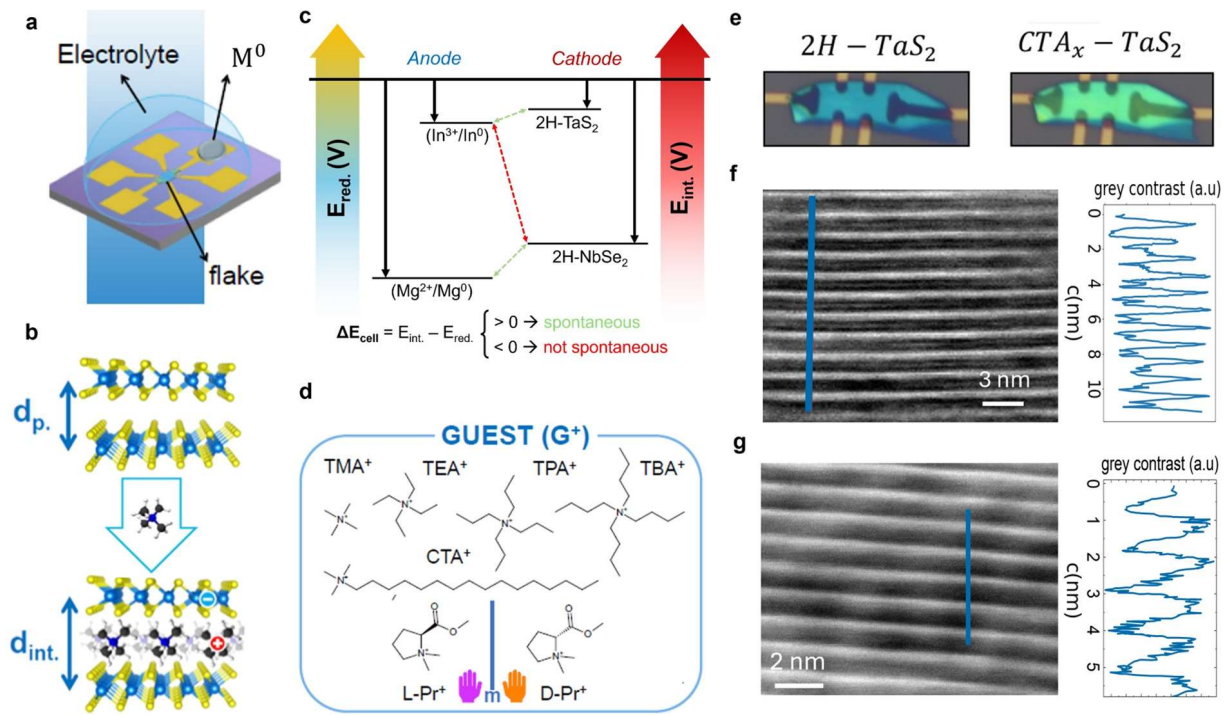

*Figure 1. In-situ galvanic molecular intercalation. a Schematic of the galvanic intercalation setup: a vdW host flake (cathode) is immersed in an electrolyte containing molecular cations and electrically connected to a reactive metal anode (M$^0$). b Charge-neutral intercalation leads*

*to an expanded interlayer distance $d_{int}$ with one electron transferred per molecular ion.* **c** *Energy diagram of the galvanic cell: intercalation proceeds spontaneously for anodic reduction potentials ($E_{red}$) lower than the intercalation potentials ($E_{int}$). $In^0$ and $Mg^0$ are used for 2H-TaS$_2$ and 2H-NbSe$_2$ intercalation, respectively.* **d** *Range of molecular guests, including linear alkylammonium series and enantiomeric chiral prolinium derivatives (L-Pr, D-Pr).* **e** *Optical micrograph of a 20nm-thick 2H-TaS$_2$ showing color change upon CTA intercalation. High-angle annular dark-field STEM images of an achiral TMA-* **f** *and a chiral D-Pr-TaS$_2$ superlattices* **g** *displaying the expected periodicity as observed in the plotted intensity profiles (right side of the STEM micrographs).*

The process yields periodic superlattices of insulating molecules and metallic TMD layers, increasing the pristine interlayer spacing $d_p$ to an intercalated distance $d_{int}$. Simultaneously, intercalation introduces a sheet carrier doping on the order of $n_{2D} \sim 10^{14} e^- cm^{-2}$, as each intercalated cation is balanced by the addition of an electron in the TMD (Figure 1b). Importantly, the entire intercalation process is performed under ambient conditions, offering a robust and scalable alternative to conventional intercalation techniques.

The influence of interlayer spacing and system symmetry on the superconducting state is investigated through the intercalation of the various molecular species depicted in Figure 1d. These include alkyl-ammonium cations with different sizes and geometries, as well as chiral L- and D-proline-derived cations, which lack both mirror and inversion symmetry (see Figures S1-S4, Supporting Information). X-ray diffraction (XRD) measurements of samples with many intercalated flakes, shown in Figure S5, Supporting Information, exhibit shifted peaks corresponding with interlayer distance $d_{int}$ in the range of 1-1.5 nm for different molecules, and no reminiscence of the parental 2H-phase. For single crystals, the uniformity of the process is confirmed through optical imaging and scanning transmission electron microscopy (STEM) observations. For thin flakes, color contrast is highly sensitive to thickness changes as shown in Figure 1e before and after CTA-intercalation of TaS$_2$. STEM measurements performed on TMA- and DPr-TaS$_2$ samples (Figure 1f) reveal the formation of a highly ordered organic/inorganic superlattices with a periodicity of approximately 1 nm, consistent with the XRD pattern.

### 2.2. Transport in hybrid superlattices.

Molecular intercalation in TaS$_2$ enhances superconductivity, as shown in Figure 2a, with achiral cations yielding a superconducting critical temperature $T_c \simeq 2.8$ K, close to the monolayer value [11], independently of the interlayer distance. The sharp transitions and minimal variation across different samples (Fig 2b), underscore both the intercalation quality and the robustness of the 2D superconductivity, particularly when compared to atomically thin samples, susceptible to surface roughness and degradation. Remarkably, chiral intercalants boost $T_c$ up to $\simeq 4.7$ K, approximately 35% above the monolayer, and exceeding values obtained using inorganic intercalants [10, 37–39]. This can be ascribed to the particular structure and chemical nature of the L-Pr$^+$ and D-Pr$^+$ ions. In particular, the carbonyl oxygen lone pairs may contribute partial charge transfer to 2H-TaS$_2$, while molecular orbitals associated with the C=O group could hybridize with states at the Fermi level, increasing the density of states and thus strengthening Cooper pairing.

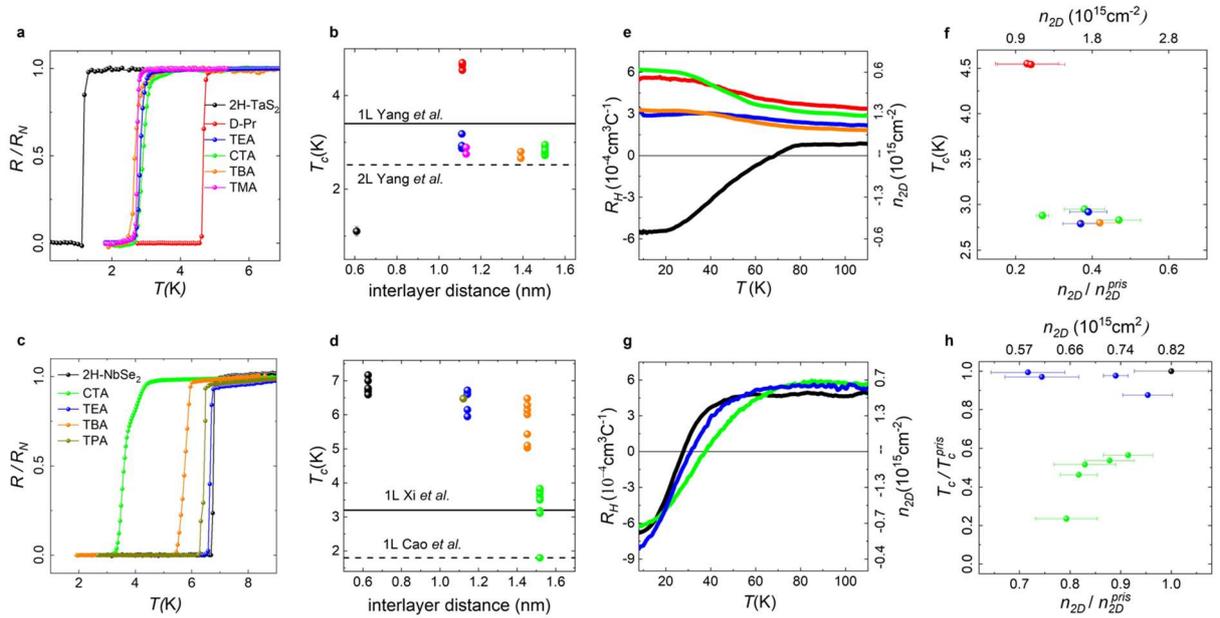

*Figure 2. Transport in hybrid superlattices. a Temperature dependence of the resistance for pristine and molecularly intercalated TaS$_2$ superlattices, showing enhanced superconductivity with $T_c \simeq 4.7K$ for chiral D-Pr-TaS2 above the monolayer limit. The resistance is normalized to the normal-state resistance $R_N$. b Extracted $T_c$ versus interlayer distance of several devices compared to 1L- and 2L-TaS2 values (solid and dashed lines) [11]. c Low-temperature dependence of the resistance for selected NbSe$_2$-based superlattices. d Critical temperature $T_c$ as a function of interlayer spacing $d_{int}$ for several NbSe$_2$ hybrid superlattices compared with reported 1L-NbSe$_2$ values [12, 40]. A pronounced dispersion in $T_c$ is found at large $d_{int}$. Temperature dependence of the Hall coefficient $R_H$ for selected TaS$_2$ superlattices. Molecular intercalation induces suppression of $R_H$ sign change and anomalous depletion of hole carriers (right axis). f Critical temperature as a function of normalized $n_{2D}$ (extracted from Hall data*

*measured at 100 K) before and after intercalation for several TaS₂ based nanodevices.* **g** $R_H(T)$ *for NbSe₂ superlattices. Sign change in $R_H$ increases with $d_{int}$ indicating a CDW enhancement.* **g** *Normalized $T_c$ as a function of $n_{2D}/n_{2D}^{pris}$ for several TEA- and CTA- NbSe₂ superlattices. The color scheme used for TaS₂ and NbSe₂ superlattices in panel* **a** *and* **c** *is applied consistently across the other panels.*

In NbSe₂, where superconductivity weakens in the monolayer limit [8, 12], intercalation effects are markedly different. As shown in Figure 2c, $T_c$ decreases monotonically with increasing intercalant size, while large intercalants display a substantial sample-to-sample variation as summarized in Figure 2d. Critical temperatures comparable to the encapsulated monolayer value (≤ 3.1 K) [12, 40] are only reached for the largest intercalant.

The impact of intercalation on the CDW order and carrier density was examined via Hall measurements. Figure 2e shows the temperature dependence of the Hall $R_H = \Delta R_{xy} t / \Delta B_\perp$ for representative TaS₂ superlattices, where $R_{xy}$, $t$ and $B_\perp$ are the transverse resistance, the flake thickness and the out-of-plane magnetic field. For pristine TaS₂, the CDW transition at $T_{CDW} \sim$ 70 K drives the sign change in $R_H$ corresponding to a carrier-type transition from hole-to electron-dominated transport [41]. Two key features emerge in the intercalated TaS₂: (*i*) suppression of the sign change, and (*ii*) a marked $R_H$ enhancement. The former indicates that hole transport persists across the full temperature range. This is associated with the CDW suppression [11] and confirmed by longitudinal and transverse transport (Figure S6, Supporting Information). On the other hand, the $R_H$ enhancement indicates that the sheet carrier density $n_{2D} = d_p / eR_H$ (right axis of Figure 2e) substantially decreases. Figure 2f plots $T_c$ as a function of the normalized carrier density $n_{2D}/n_{2D}^{pris}$ measured at T=100 K for several devices, where $n_{2D}^{pris}$ is the value obtained before intercalation [42]. The hole concentration decreases by approximately $10^{15} cm^{-2}$. Notably, the recorded change in charge carrier density is highest in the case of L-Pr and D-Pr, confirming that the highest $T_c$ is related to a larger charge doping. While the decreased hole density is qualitatively consistent with the electron doping introduced by intercalation, the reduction exceeds the nominal charge transfer of one electron per intercalated molecule by about an order of magnitude. To account for this discrepancy, we propose that the intercalated molecules effectively separate the layers, restricting transport to only a few of them. As a result, the hole density probed by R_H is reduced not only by electron doping, but also by the smaller number of layers effectively contributing to transport. The hypothesis is supported by transport measurements in partially intercalated Pr-TaS$_2$ (Figure S7, Supporting Information) in which $T_c$ decreases, dropping the hole carrier density.

By contrast, Hall measurements in NbSe$_2$ show moderate intercalant effects. For TEA- and CTA-based devices, representing the smallest and largest cations, the sign change in $R_H$ associated with the CDW transition, shifts to higher temperatures in line with the power-law trend of the longitudinal resistivity (Figure S8, Supporting Information). Figure 2h shows the extracted sheet carrier densities over several samples, with a charge depletion lower than $3 \cdot 10^{14} cm^{-2}$, consistent with one-to-one molecular charge transfer. This indicates that, unlike in TaS$_2$, in NbSe$_2$ the charge transport occurs across the whole crystal thickness. Importantly, weak effects in the superconducting order are found for TEA-based devices as previously reported [30]. Instead, CTA-intercalated superlattices show a correlation between charge depletion and $T_c$, suggesting that a minimum interlayer decoupling is required before charge transfer significantly influences superconductivity.

**2.3. Ising superconductivity in hybrid superlattices.**

We next examine the macroscopic manifestations of SOC in hybrid superlattices. Figure 3a presents the angular anisotropy of the upper critical field $H_{c2}$ given by $\gamma(\theta) = H_{c2}(\theta)/H_{c2}^\perp$ for TEA-TaS$_2$ superlattice measured at T = $0.8T_c$. The data are well captured by a 2D Ising superconducting model (see methods), with an in-plane critical field at $\theta = 0$ exceeding the Pauli paramagnetic limit $H_P = 1.84 k_B T_C$ at the given temperature (horizontal line).

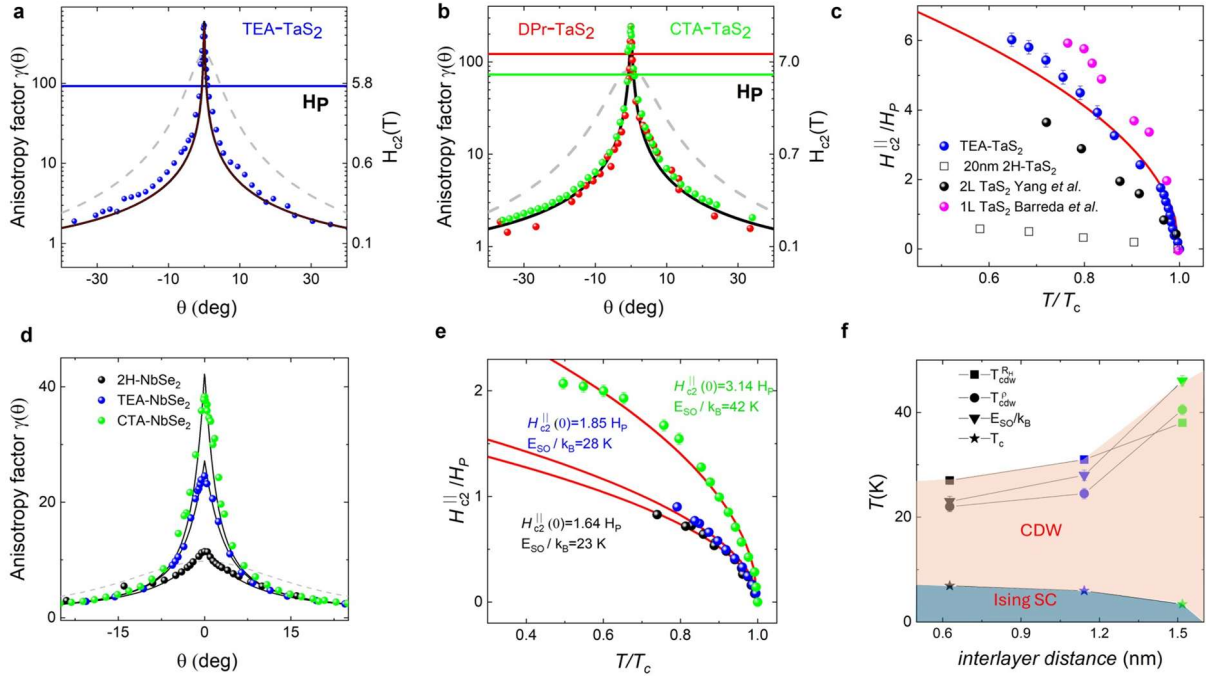
9

*Figure 3. Ising superconductivity in hybrid superlattices.* **a** *Angular anisotropy of the upper critical field* $\gamma(\theta) = H_{c2}(\theta)/H_{c2}^{\perp}(0)$ *for TEA-TaS$_2$ superlattice. Data fit well to a 2D anisotropic model (black line), in contrast to a 3D anisotropic bulk behavior (dashed grey). The Pauli limit $H_P = 1.84T_c$ is shown by the horizontal line.* **b** *Ising superconductivity for chiral and CTA intercalated TaS$_2$ superlattices.* **c** *Normalized $H_{c2}(T)/H_P$ for TEA-intercalated (blue dots), 20 nm-thick 2H-TaS$_2$ (empty squares), 1L- and 2L-TaS$_2$ [11,13]. $\gamma(\theta)$* **d** *and $H_{c2}(T)/H_P$* **e** *for selected NbSe$_2$-based superlattices. Black lines correspond to the 2D model fits in* **d** *$H_{c2}^{\parallel}(0)$. Spin-orbit temperatures $\Delta_{SO}/k_B$ are extracted from Tinkham fits (red lines).* **f** *Phase diagram for NbSe$_2$ superlattices showing a correlation between the spin-orbit temperature and $T_{CDW}$ (extracted from transverse and longitudinal transport).*

For comparison, we also fit the curves with a conventional 3D model (dashed grey lines), which fails to reproduce the sharp, cusp-like anisotropy near θ = 0. The 2D model accounts for the observed cusp and large enhancement of the critical field independently of the molecule as observed for chiral D-Pr and CTA-intercalated superlattices (Figure 3b). To confirm monolayer behavior, Figure 3c shows the normalized $H_{c2}^{\parallel}(\theta = 0)$ for a representative hybrid superlattice, a 20 nm-thick pristine sample, 1L and 2L TaS$_2$ (Refs. [11, 13]). Ising enhancement exceeds the reported for the bilayer, confirming the effective interlayer decoupling upon intercalation with a zero-temperature critical field $H_{c2}^{\parallel}(0) \sim 7H_p = 32T$ extracted from the Tinkham fit (red line). Although weaker, Ising physics is also evident in NbSe$_2$ superlattices with the characteristic cusp-like $H_{c2}(\theta)$ as shown in Figure 3d. Notably, $H_{c2}^{\parallel}$ increases with the intercalant size, indicating progressive interlayer decoupling. This further confirms that the charge transport in intercalated NbSe$_2$ retains a more 3D character as compared to TaS$_2$. We quantify the Ising enhancement in NbSe$_2$ superlattices by extracting $H_{c2}^{\parallel}(0)$ from the temperature dependence shown in Figure 3e. The spin-orbit field $B_{SO}$ and the spin-splitting energy $\Delta_{SO} = \mu_B B_{SO}$ are obtained from the modified Zeeman relation $1/2(H_{c2}^2/B_{SO})\mu_B = 1/2\mu_B H_P$ [12]. Remarkably, $\Delta_{SO}$ in NbSe$_2$ superlattices reaches values comparable to the CDW energy $k_B T_{CDW}$, obtained from longitudinal and transverse transport. This underscores the cooperative interplay between Ising superconductivity and the density-wave state, as well as their tunability with the interlayer distance, summarized by the phase diagram in Figure 3f.

**2.4. Dissipationless transport and superconducting symmetries.**

We next turn to dissipationless transport in monolayer-like TaS$_2$ superlattices and the underlying symmetries. Figure 4a shows that the critical current $I_c$ increases by two orders of magnitude upon intercalation, consistent with the enhancement observed in the 2D limit [7].

Figure 4b shows the temperature dependence of the normalized critical current density $J_c(t)/J_c(0)$, where $t = T/T_c$ is the reduced temperature. All devices collapse onto an universal curve, demonstrating that intercalation leaves the temperature scaling unchanged. The data are accurately described by the phenomenological form $J_c(t)/J_c(0) = (1-t^2)^\alpha (1+t^2)^\beta$, commonly used to identify vortex-pinning mechanisms [43-46]. The extracted exponents, $\alpha = 5/6$ and $\beta = 2/3$, differ markedly from those expected for disorder-induced $\delta T_c$ pinning ($\alpha = 7/6, \beta = 5/6$), mean-free-path fluctuations, $\delta l (\alpha = 5/2, \beta = -1/2)$ or the Bardeen decay in dirty superconductors [47]. Instead, $J_c$ remains robust under thermal depairing, indicative of the presence of strongly correlated electrons, akin to non-phonon superconductors near a quantum critical point [43, 48, 49]. This positions multiband TMDs as strong candidates for realizing unconventional superconductivity.

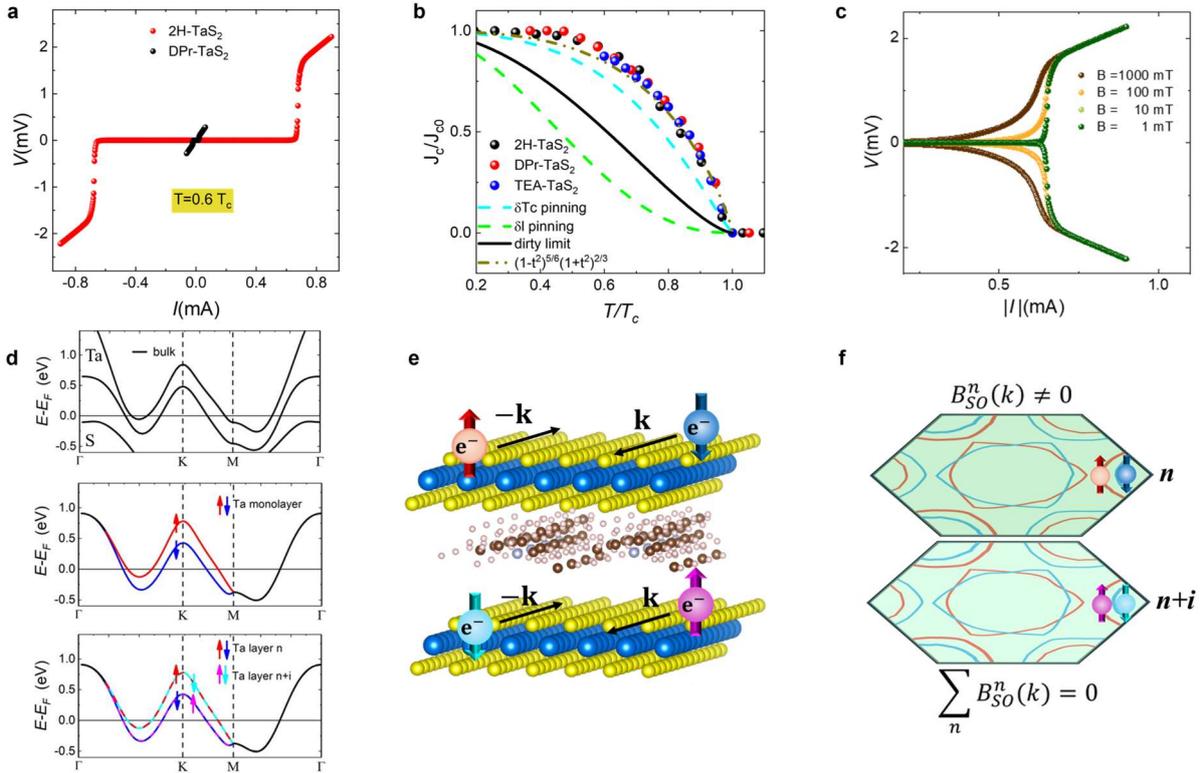

*Figure 4. Dissipationless transport and electronic structure in TaS$_2$ superlattices. a Current-voltage characteristics for 2H-TaS$_2$ before and after D-Pr-intercalation taken at T=0.6T$_c$. b Temperature dependence of the normalized critical density of currents for pristine and given intercalated systems fitted to a two-exponent model $(1-t^2)^\alpha (1+t^2)^\beta$. δTc, δl -pining mechanisms and Bardeen fitting of dirty superconductors fail to describe the trend, well described by $\alpha = 5/2$ and $\beta = 6/5$ as observed in unconventional superconductors [46]. c Reciprocal disspationless transport under selected B$_\perp$ d DFT-calculated band structures for bulk, monolayer and intercalated ($d_{int} = 1.1nm$) TaS$_2$, along the $K - \Gamma - M$ path. Red and*

*blue arrows indicate Ta spin-split bands due to SOC. In the intercalated system, degenerate bands from adjacent Ta layers exhibit opposite spin polarization (cyan and purple arrows), yielding no net spin polarization.* ***e*** *Schematic of spin-layer locking in intercalated structures.* ***f*** *Fermi surfaces of adjacent layers with opposite spin-splitting bands and no net spin polarization as obtained in* ***d***.

Nonreciprocal dissipationless transport where $I_c$ depends on the bias polarity (+/-) provides a sensitive probe of broken symmetries. The so-called superconducting diode effect requires simultaneous breaking of time-reversal and inversion symmetry [47, 50-55]. The latter is broken in noncentrosymmetric 1H-TaS$_2$, as do the unconventional pairing states proposed for the monolayer [18, 19], but dissipationless transport under time-reversal-breaking magnetic fields remains strictly reciprocal ($I_c^+ = |I_c^-|$, Figure 4c) for all the monolayer-like superlattices investigated. DFT calculations shed light on these contradictory results. The band structure for bulk 2H-TaS$_2$, 1H-TaS$_2$, and an intercalated structure with $d_{int}$=1.1 nm and carrier doping of $2 \cdot 10^{14} e^-/cm^2$ are represented in Figure 4d. In monolayers, lack of perpendicular dispersion results in degenerated Ta bands while the S band is pushed away from the Fermi level. At the K valleys, Ising SOC induces an out-of-plane effective magnetic field $\boldsymbol{B_{SO}}(\boldsymbol{k})$ producing Ta spin-splitting bands (red and blue lines). In the intercalated structure, interlayer decoupling yields a monolayer-like band structure with degenerate Ta bands, with negligible contribution of S $4p_z$ orbitals to the density of states (Figure S10, Supporting Information) and interlayer hopping $\tau_\perp = 0$. However, no spin polarization emerges at $\boldsymbol{K}$, since adjacent, degenerate layers carry opposite spin splitting bands. Thus, inversion symmetry is broken locally at the layer level $\boldsymbol{B_{SO}^n}(\boldsymbol{k}) = -\boldsymbol{B_{SO}^{n+1}}(\boldsymbol{k}) \neq 0$ while preserved globally $\sum_n \boldsymbol{B_{SO}^n}(\boldsymbol{k}) = 0$ as illustrated in Figure 4e,f. Additional calculations in Figure S11, Supporting Information identify $d_{int} = 0.9 nm$ as the threshold for lifting degeneracy. Hall measurements and electronic calculations are consistent with a framework of quasi-2D hybrid superlattices acting as electronically decoupled monolayers with a few-layer transport thermally activated. Below $T_c$, spin-momentum locking within each layer stabilizes monolayer Ising superconductivity, while Josephson coupling between oppositely spin-polarized bands across adjacent layers ensures reciprocal transport. In contrast, NbSe$_2$ superlattices preserve bulk-like properties: Hall transport probe the entire crystal and small intercalants induce negligible modifications to the superconducting order, consistent with previous bulk studies [30]. This dichotomy originates from differences in orbital character. Although both TMDs share similar band topologies, in NbSe$_2$ dispersive 3D bands from Se $p_z$ orbitals cross the Fermi level. This results in a larger effective interlayer hopping $\tau_\perp$ compared with TaS$_2$, consistent with previous estimations of $\tau_\perp =$10 and 20 meV for TaS$_2$ and NbSe$_2$, respectively [13]. This difference can be intuitively ascribed to the 10% smaller radial extent of $p_z$ orbitals in S compared to their Se counterparts. Given the exponential dependence of $\tau_\perp$ on the interlayer distance, the larger interlayer coupling in NbSe$_2$ allows it to maintain its three-dimensional character upon intercalation, whereas TaS$_2$ undergoes a crossover to a more two-dimensional electronic structure.

## 3. Conclusion

In summary, we have investigated organic-intercalated, nanometer-thick TaS$_2$ and NbSe$_2$ with contrasting thickness-dependent electronic states. In TaS$_2$, intercalation induces interlayer decoupling and monolayer-like transport, whereas NbSe$_2$ retains its 3D character upon intercalation, requiring significant interlayer decoupling before charge-transfer tunes the superconducting order. Superlattices exhibit reciprocal dissipationless transport, indicative that chirality alone does not induce unconventional pairing and ruling out parity-breaking states reported in the 2D limit. Instead, the behavior is consistent with electronically decoupled monolayers whose opposite spin-splitting bands restore inversion symmetry. These findings disentangle the roles of interlayer coupling and molecular charge transfer in hybrid TMD superconductors, and establish molecule-intercalated compounds as a robust, device-integrated platform to engineer and probe correlated superconducting states.

## 4. Methods

**Sample fabrication.** Heterostructure devices are fabricated using a dry-transfer method. 2H-NbSe$_2$ and 2H-TaS$_2$ flakes were mechanically exfoliated onto polydimethyl- siloxane (PDMS) stamps. Nanometric-thick flakes were identified optically and transferred onto pre-patterned Ti (2 nm)/Au(8 nm) electrodes on Si/SiO$_2$ substrate. Galvanic intercalation is conducted under ambient conditions by contacting a selected metal anode $M^0$ with low reduction potential to one of the electrode pads. The galvanic cell with the van der Waals flake acting as the cathode is immersed in a non-aqueous electrolyte containing a salt of the target molecule. Spontaneous oxidation of the metal drives the re- duction of the flake. To preserve charge neutrality, molecular cations are inserted into the van der Waals gap.

**Structural characterization.** X-ray diffraction was performed using an Empyrean diffractometer (PANalytical) equipped with a copper cathode source. Both K$\alpha$1 ($\lambda$=1.5406 Å) $K\alpha$2 ($\lambda$=1.5443 Å) lines were employed to maximize signal intensity. Measurements were taken on both bulk crystals and exfoliated flakes deposited on Au(30–50 nm)/Ti(3 nm)/SiO$_2$ substrates. For single crystals, intercalation was first verified by comparing thickness of pristine and intercalated flakes using atomic force microscopy (AFM 5500 Agilent), which also provided color-contrast calibration for optical images. Flake thickness in devices was estimated optically before and after intercalation to limit air exposure times to under 1 hour, with final AFM measurement used for resistivity calculations. High-angle annular dark-field (HAADF)-scanning transmission electron microscopy (STEM) measurements were developed using an aberration probe-corrected Thermo Fisher Scientific Titan Low-Base microscope operated at 300 keV. A Thermo Fisher Scientific Helios 650 focused ion beam (FIB) instrument was employed for TEM lamella preparation.

**Transport measurements.** Electrical transport was measured in Quantum Design Physical Property Measurement Systems (PPMS). Pre-patterned contacts were designed for both longitudinal and Hall configurations. To minimize artefacts from carrier-density variations, $n_{pris}$ was measured before and after intercalation and normalized to the average value. The 2D and the 3D models for the angular anisotropy of the upper critical field $H_{c2}$ are obtained from the Tinkham formula $H_{c2}^{\parallel} \propto \left(1 - \frac{T}{T_c}\right)^{0.5}$ and $H_{c2}^{\parallel} \propto \left(1 - \frac{T}{T_c}\right)$, respectively. DC current-voltage characteristics were obtained by sweeping a low-noise current in both polarities and recording the voltage drop across the flakes.

**DFT calculations.** The first-principles calculations were performed using spin-polarized density-functional theory (DFT) in the SIESTA code implementation [56] in its 5.0.0-beta1 version. A double-ζ polarized (DZP) basis was employed for all atoms, generated with a 50 meV energy shift and a split norm of 0.15 [56]. Core electrons were replaced by fully relativistic norm-conserving pseudopotentials [57] in PSML format generated using the ONCVPSP method [58] and obtained from the PseudoDojo repository [59] available at the web interface http://pseudo-dojo.org/, which include both scalar relativistic effects and spin-orbit coupling. For Ta, the valence configuration includes 5s2 5p6 5d3 6s2, with cutoff radii of 1.70 Bohr for s, p, and d, and for S, the valence configuration is 3s2 3p4, using cutoff radii of 1.45 Bohr for s and p. The exchange-correlation functional was approximated using the generalized-gradient approximation (GGA) as parametrized by Perdew-Burke-Ernzerhof (PBE) [60]. A real-space mesh cutoff of 300 Ry and a tolerance of $10^{-6}$ eV for the density matrix were used for the convergence of the self-consistent field (SCF). To eliminate the artificial interaction between periodic replicas, a vacuum space of 22 Å was added along the out-of-plane direction. Full spin-orbit coupling (SOC) was included, and the Brillouin zone was sampled by using a Monkhorst-Pack grid [61] of 24x24x6 for the bilayer and 24x24x1 for the monolayer system.

**Acknowledgements**

This work was supported under Projects PID2021-128004NB-C21, PID2021-122511OB-I00, PID2024-157558NB-C22, and PID2024-155708OB-I00 and PID2022-139776NB-C65 funded by Spanish MICIU/AEI/10.13039/501100011033 and by ERDF A way of making Europe; and under the María de Maeztu Units of Excellence Programme (Grant CEX2020-001038-M). It is also supported by MICIU/AEI and by the European Union NextGeneration EU Plan (PRTR-C17.I1), by the IKUR-Quantum Technologies Strategy under the collaboration agreement between DIPC, CFM and CIC nanoGUNE on behalf of the Department of Education of the Basque Government, FPI grant PREP2022-000453 and María de Maeztu award to Nanogune, Grant CEX2020-001038-M. M.G. acknowledges support from MICIU/AEI and European Union NextGeneration EU/PRTR (grant no. RYC2021-031705-I). I.R. acknowledges support from Project PID2023-151549NB, funded by MICIU/AEI and by FEDER, UE; Basque Government IT1553-22. R.A. acknowledges funding from the Spanish MICIU (PID2023-151080NB-I00/AEI/10.13039/501100011033), from the Government of Aragon (project DGA E13-23R), from MICIU with funding from EU Next-Generation (PRTR-C17.I1) promoted by the DGA and support from the 'Severo Ochoa' Programme for Centres of 758 Excellence in R&D of the Spanish MICIU (CEX2023-759 001286-S MICIU/AEI/). TEM studies have been conducted in the Laboratorio de Microscopias Avanzadas (LMA) at Universidad de Zaragoza. E.A. also acknowledges support from the United Kingdom's EPSRC Grant no. EP/V062654/1. The authors acknowledge the technical and human support provided by the DIPC Supercomputing Center for the computational resources.

**Data Availability Statement**

The data of this study are available from the corresponding author upon reasonable request.

Supporting Information

**Degenerate monolayer Ising superconductors via chiral-achiral molecule intercalation in devices**

*Daniel Margineda\*, Covadonga Álvarez-García, Daniel Tezze, Sanaz Gerivani, Mohammad Furqan, Iván Rivilla, Fèlix Casanova, Raul Arenal, Emilio Artacho, Luis E. Hueso, and Marco Gobbi\**

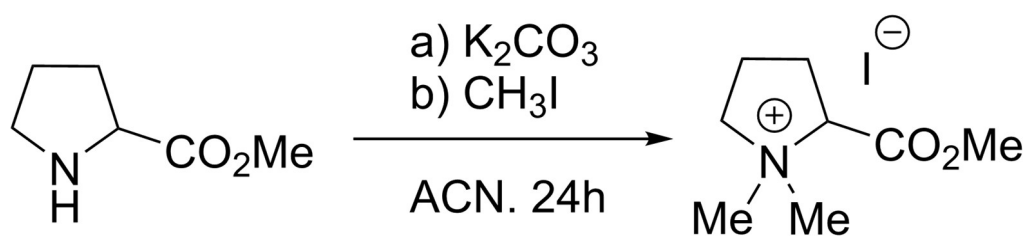

**Figure S1. General Synthesis of (S)- and (R)-2-(methoxycarbonyl)-1,1 dimethylpyrrolidin-1-ium iodide (L-PrI and D-PrI).** Schematic of the synthesis of L-Pr and D-Pr proline derivatives in acetonitrile (ACN) solvent.

SYNTHESIS AND CHARACTERIZATION OF CHIRAL PROLINE-BASED MOLECULES.

To a solution of N-methyl-(D or L)-proline methyl ester (1.0 mmol) in CH3CN (150 mL), K2CO3 (1.0 mmol) was added. The reaction mixture was stirred at room temperature for 3 h, followed by the dropwise addition of CH3I (17.5 mL). Stirring continued for 24 h. The resulting precipitate was collected by filtration. The filtrate was concentrated under reduced pressure to yield additional product. Recrystallization from CHCl3 afforded N,N-dimethyl-(D or L)-proline methyl ester iodide in high yield. Reaction scheme, NMR, g-COSY, and HSQC spectra are shown in Fig. 1, 2, 3, respectively. [1–3]

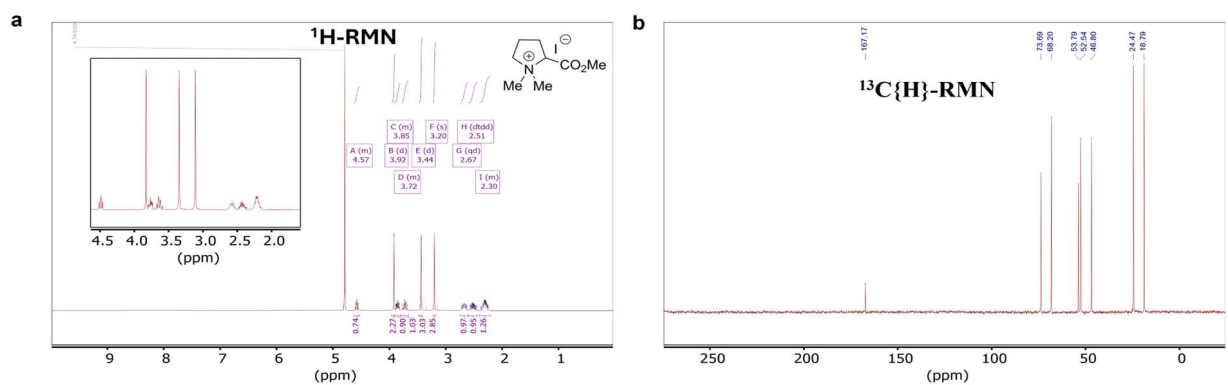

**Figure S2. Nuclear Magnetic Resonance Spectroscopy. a** 1H NMR and **b** 13C NMR spectra were recorded at 400 and 101 for 13C NMR, equipped with a z gradient BBOF probe, in CDCl3. The data are reported as s = singlet, d = doublet, t = triplet, q = quartet, p = quintet, m = multiplet or unresolved, br s = broad signal, coupling constant(s) in Hz, integration. The 1H spectra were recorded using noesygppr1d sequence from Bruker's library at 400 MHz. A time domain of 64 k and a spectral width of 10000 Hz. Interpulse delay: 1 s. Acquisition time: 3 s. Number of scans: 64. Mixing time: 0.01 s. Only the spectra for D-Pr+ are shown, since solution NMR spectroscopy is not able to distinguish between enantiomers and therefore the spectra for both compounds D and L are identical.

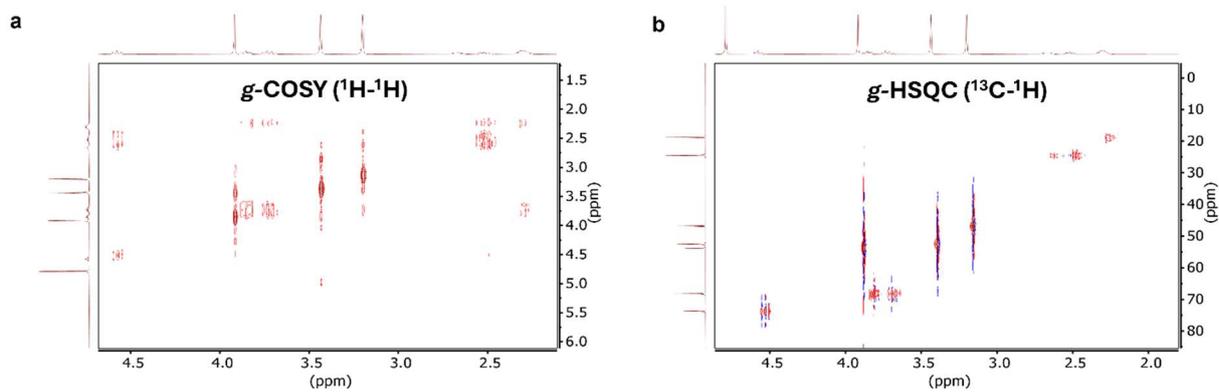

**Figure S3. Organic correlations in D-Pr based solutions. a** Correlation spectroscopy (g-COSY) and **b** Heteronuclear Single Quantum Coherence (HSQC) showing protonic and proton-carbon correlation in D-Pr based solutions.

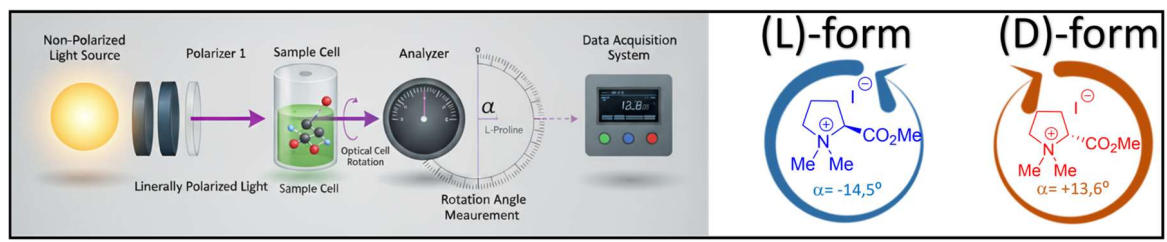

**Figure S4. Optical activity of L-Pr and D-Pr-MeOH solutions**. Negative (–14.5°) and positive (+13.6°) rotations of the incident light's polarization are observed in L-Pr- and D-Pr-based solutions, respectively, consistent with the presence of left-handed and right-handed chiral molecules.

Optical activity of proline derivatives was measured using an equipment JASCO P-2000 digital polarimeter, which allows for high-precision determination of molecular chirality through the rotation of plane-polarized light. This method operates by exploiting the interaction between chiral molecules and polarized light, producing a detectable rotation in the plane of polarization. L-Pr and D-Pr chiral molecules were dissolved in MeOH with a concentration of 0.128 w/v and loaded into a thermally jacketed cylindrical quartz cell with dimensions (10 x 100 190-2700 nm) and a path length l= 100 mm. Optical rotations of the samples were measured using the Sodium line ($\lambda$= 589 nm with an aperture of A= 8.0 mm in a temperature-controlled chamber T=25∘ C. A negative (positive) angle rotation is observed for L-Pr (D-Pr) solutions, respectively as shown in Fig. 4, confirming enantiomeric purity.

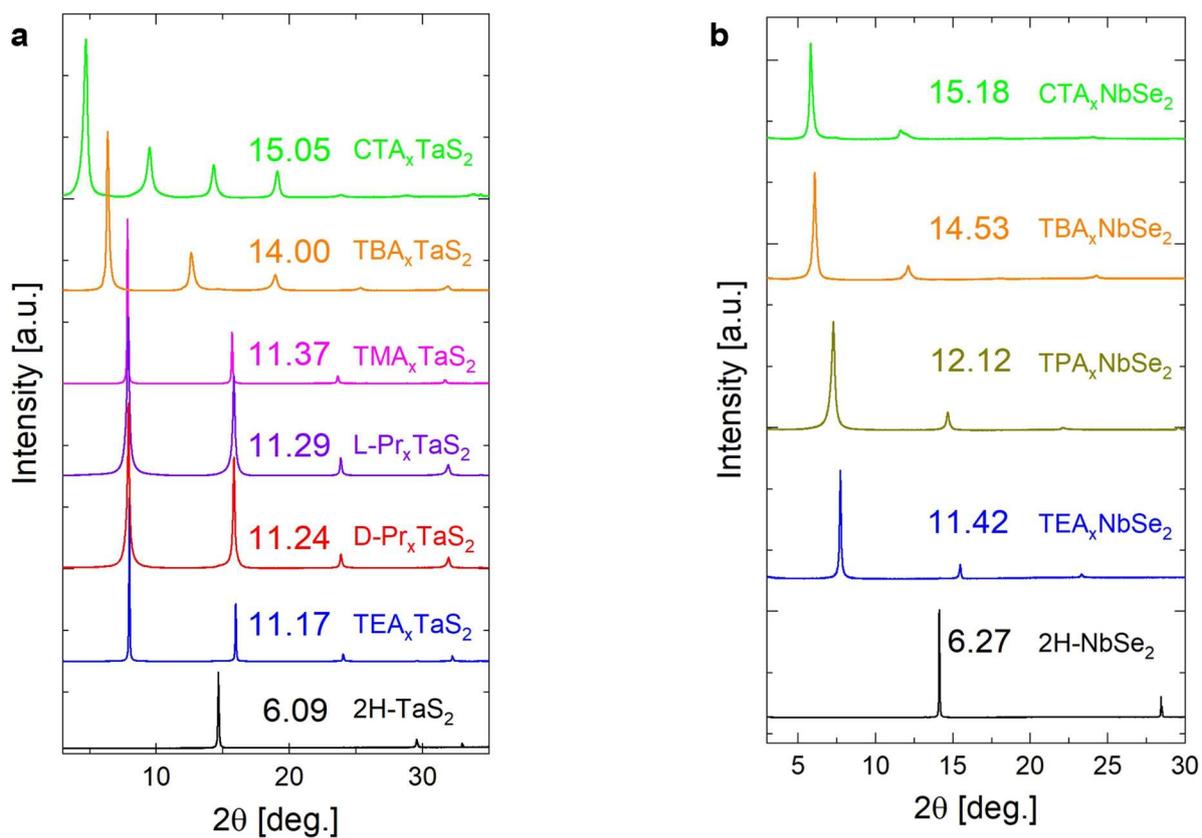

**Figure S5. X-Ray diffraction (XRD) of hybrid superlattices.** XRD measurements of molecular-intercalated TaS$_2$ **a** and NbSe$_2$ **b** obtained from exfoliated flakes deposited on Au/Ti/SiO$_2$ substrates. Interlayer distances extracted from the low-angle peak are given in Å.

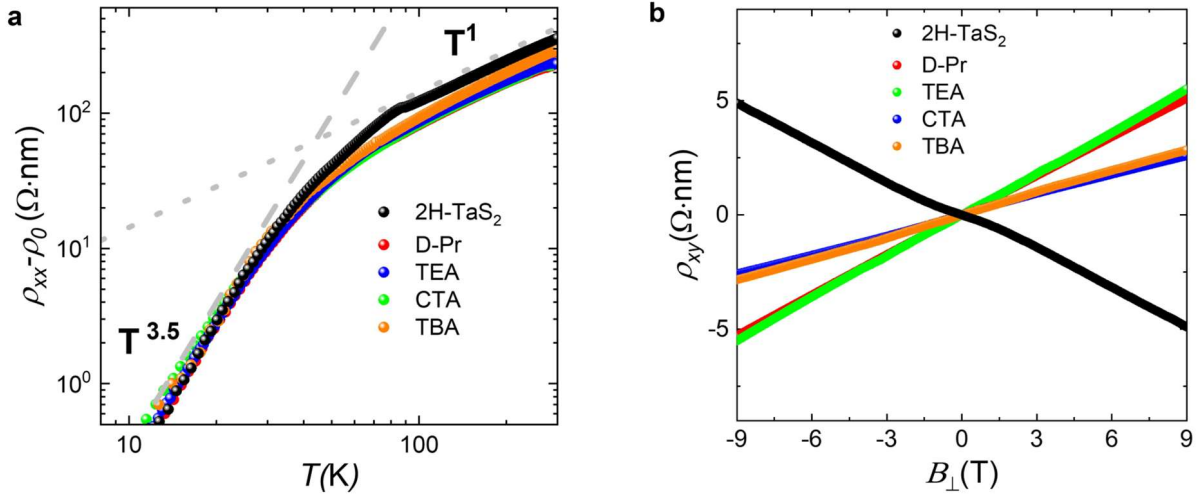

**Figure S6. Transport in TaS₂ superlattices. a** Logarithmic resistivity above $T_c$ with suppressed CDW-associated upturn at $T_{CDW} \simeq 70K$ for intercalated devices. Dashed and dot grey lines are fits to $\propto T^\alpha$ below and above $T_{CDW}$. Negligible effects to the power-law are introduced by intercalation with α ~ 3.5, an intermediate regime between electron-electron α ~ 2 and electron-phonon α ~ 5 scattering that we associated with phonon-assisted interband scattering. **b** Transverse magnetoresistivity $\rho_{xy}(B)$ at T=10 K, with slope reversal upon intercalation due to electronic band reconfiguration.

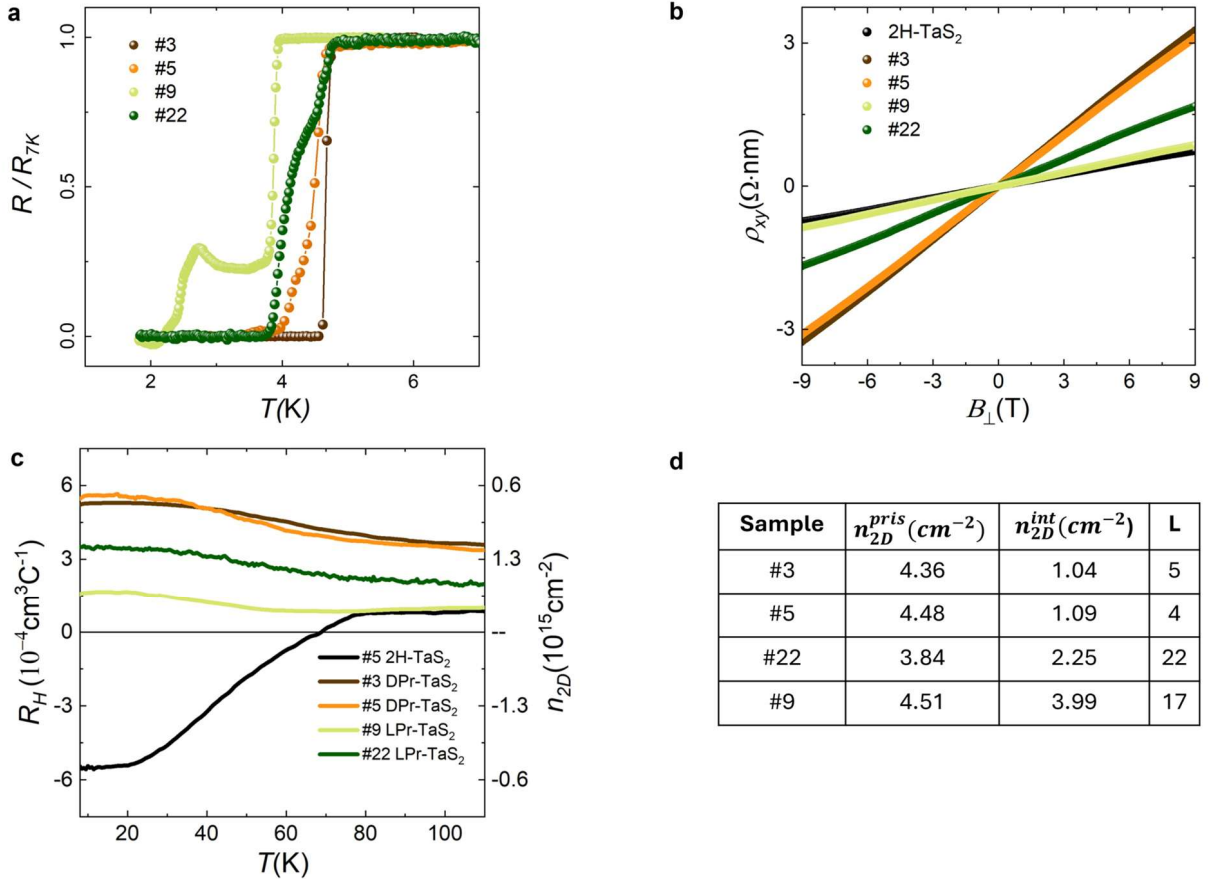

**Figure S7. Transport in partial intercalated Pr-TaS$_2$ systems. a** Low-temperature dependence of the normalized resistance for total (#3,#5) and partial intercalated (#9,#22) Pr-TaS$_2$ samples. **b** Low-temperature transverse magnetoresistance **c** and temperature dependence of the Hall coefficient for samples in **a**. Sheet carrier density extracted before and after intercalation and the estimated number of layers *L* driven transport at T=100 K. *L* is calculated assuming a molecular charge transfer of $2 \cdot 10^{-14} e^-/cm^{-2}$

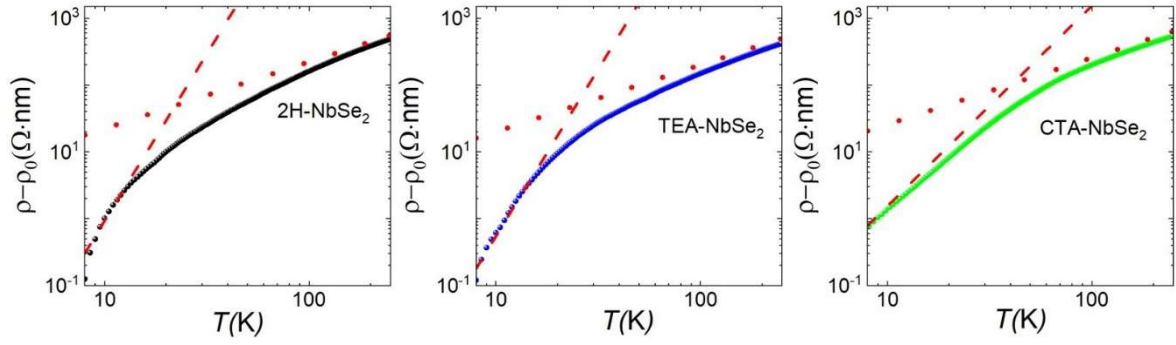

**Figure S8. Transport in NbSe$_2$ superlattices.** Logarithmic resistivity with the change in the power-law dependence ascribed to the CDW transition highlighted by the red lines.

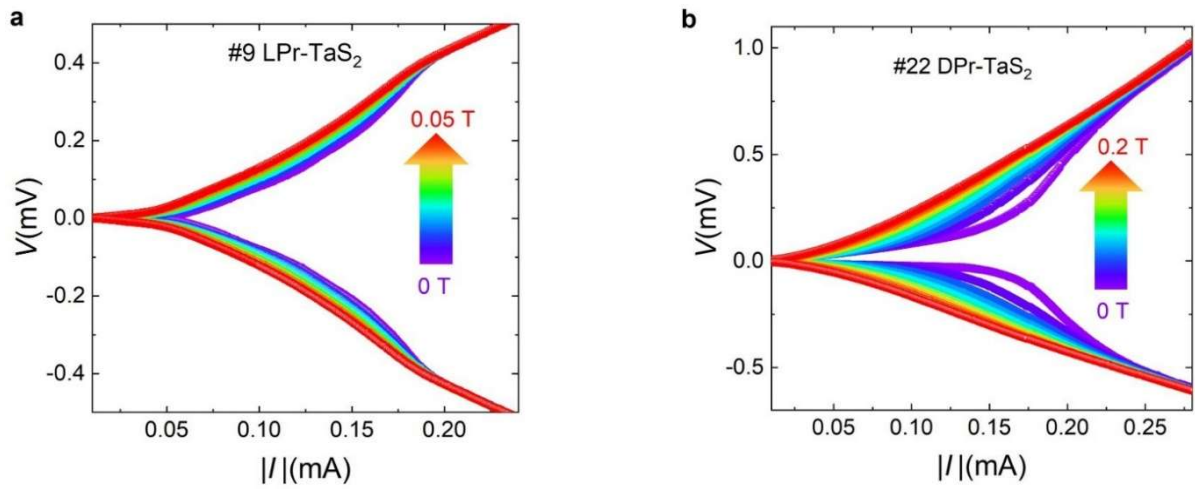

**Figure S9. Recirpocal dissipationless transport in partial intercalated Pr-TaS$_2$ systems.** IV curves under characteristics out-of-plane magnetic fields for sample #9 (a) and #22 (b) from Fig. S7.

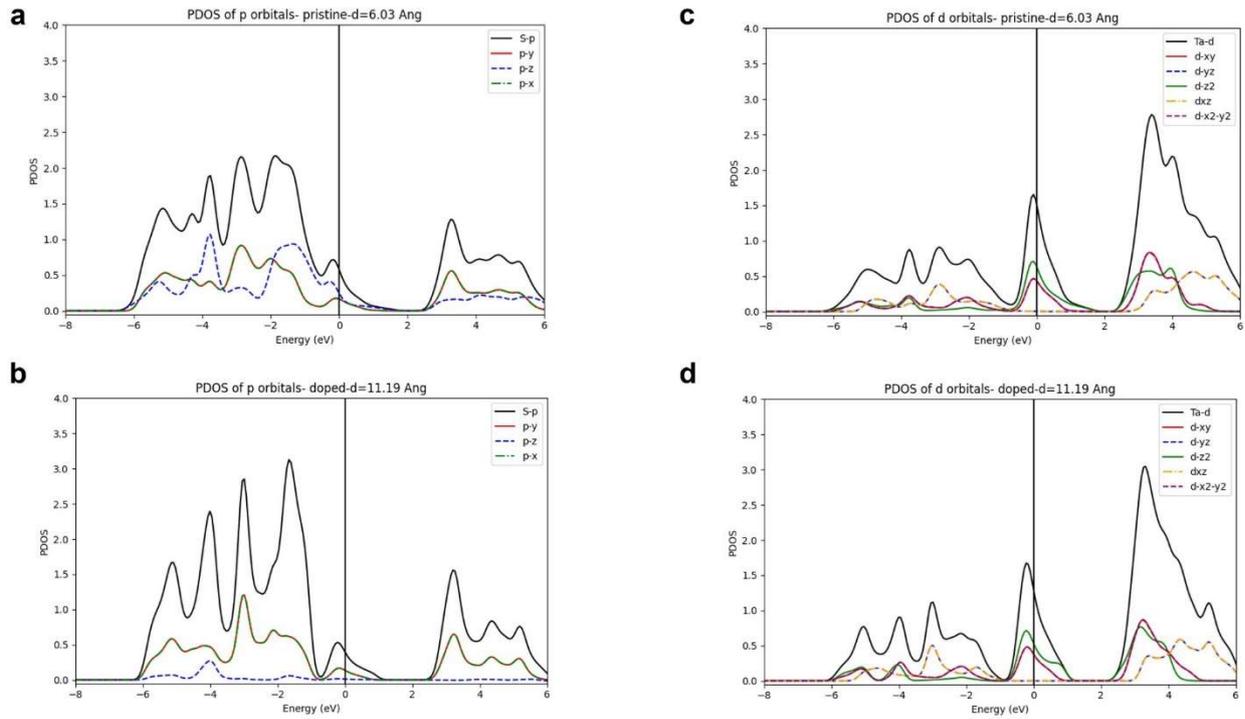

**Figure S10. Partial density of states.** Partial DOS around the Fermi level of *p* and *d* orbitals for **(a,c)** pristine and **(b,d)** intercalated samples ($d_{int}$ =1.1 nm). DOS contribution of the S $p_z$ orbitals dramatically decrease for the given interlayer distance.

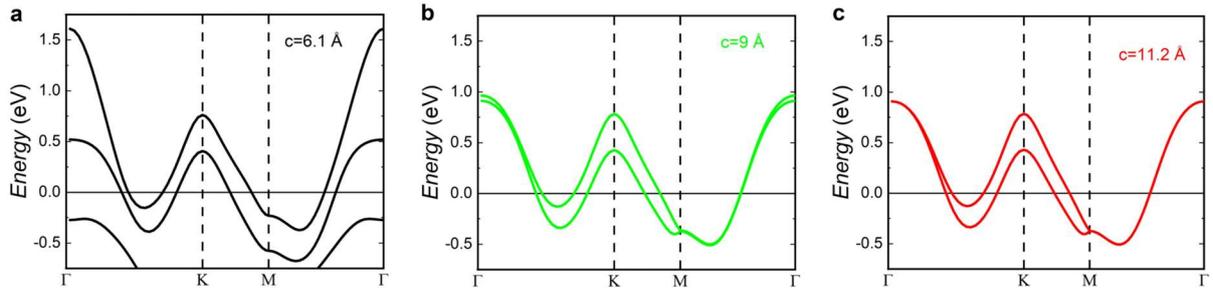

**Figure S11. Electronic structure of doped TaS$_2$**. DFT-calculated band structures for doped ($2 \cdot 10^{-14} e^- cm^{-2}$) TaS$_2$ for interlayer distance $d_{int}$=0.61 **a**, 0.9 **b**, and 1.1 nm **c**. Ta *d* bands around Γ becomes degenerated for interlayer distances larger than 0.9 nm.

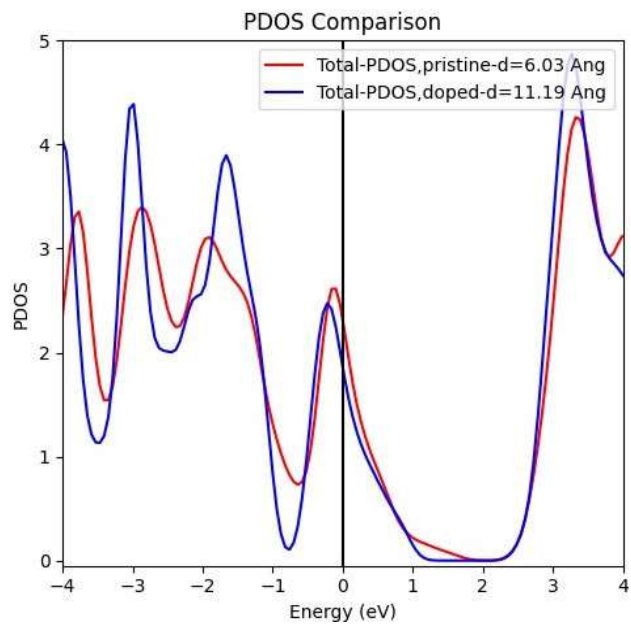

**Figure S12 Total density of states.** Total DOS around the Fermi level for pristine (red line) and intercalated ($d_{int}$ =1.1 nm) samples (blue line). Van-Hove singularity arises just below the Fermi level (Energy= 0 eV).

**References**

(1) H.-S. Gao, Z.-G. Hu, J.-J. Wang, Z.-F. Qiu, and F.-Q. Fan, Synthesis and Properties of Novel Chiral Ionic Liquids from l-Proline, Aust. J.Chem. 61, 521 (2008).

(2) J.-W. Shin, Optical activity of trans-4-hydroxy-l-proline in sodium chloride solutions, Chem. Phys. Lett. 838, 141086 (2024).

(3) C. Hongo, M. Shibazaki, S. Yamada, and I. Chibata, Preparation of optically active proline. Optical resolution of N-acyl-Dl-proline by preferential crystallization procedure, J. Agric. Food Chem. 24, 903 (1976).